\documentclass[a4paper,11pt]{article}
\pdfoutput=1 

\usepackage{jcappub} 

\usepackage[T1]{fontenc} 

\usepackage[normalem]{ulem}

\usepackage{soul} 
\usepackage{graphicx}
\usepackage{multicol,multirow}
\usepackage{amsmath,amssymb,amsfonts}
\usepackage{mathrsfs}

\usepackage{rotating}
\usepackage{appendix}

\title{\boldmath 
Towards the Reconstruction of a Unified
Dark Matter Halo: a Phenomenological Approach
}

\author[a]{Claudia Caputo,}
\author[b]{Daniele Bertacca,}
\author[c,d]{Alberto Bassi,}
\author[b]{Sabino Matarrese}

\affiliation[a]{Institute of Theoretical Physics, Faculty of Mathematics and Physics,
Charles University,\\ CZ-180 00 Prague, Czech Republic}
\affiliation[b]{Department of Physics ”Galileo Galilei”, University of Padova, INFN (Padova),\\ via F. Marzolo, 8 I-35131 Padova, Italy}
\affiliation[c]{Department of Physics, ETH Zurich, Zurich, Switzerland}
\affiliation[d]{SIAM Department, Eawag (ETH), Dübendorf, Switzerland}

\emailAdd{claudia.caputo@matfyz.cuni.cz}
\emailAdd{daniele.bertacca@pd.infn.it}
\emailAdd{abassi@student.ethz.ch}
\emailAdd{sabino.matarrese@pd.infn.it}

\abstract{
We investigate static, spherically symmetric halo configurations within 
Unified Dark Matter (UDM) scalar-field models, developing a systematic 
mapping between standard cold dark matter (CDM) density profiles and their 
UDM counterparts. Exploiting the equivalence-class structure of UDM models, 
we show that, in principle, different Lagrangian realisations can share the 
same weak-field rotation curve while exhibiting distinct field properties.
We reconstruct the effective energy density, radial and tangential pressures 
from a phenomenological circular velocity profile, ensuring the absence of 
ghosts and instabilities and the preservation of the Null Energy Condition (NEC).
Applying our procedure to several commonly used CDM halo profiles --- including 
Persic, Salucci \& Stel, NFW, and Burkert models --- we demonstrate that their 
phenomenological success can be retained within a relativistic UDM framework, 
reproducing the observed flatness of rotation curves without introducing separate 
dark matter and dark energy components.
}

\usepackage{xcolor}

\newcommand{\ud}{\mathrm{d}}

\begin{document}
\maketitle
\flushbottom
\nocite{*}

\section[Introduction]{Introduction}
\label{sec:introduction}

One of the most persistent open problems in modern physics concerns the nature of the dark components of the Universe.
Evidence for the existence of Dark Matter (DM) first emerged from gravitational dynamics on galactic scales. 
Galaxies within clusters move at speeds too high for their gravitational hold to be explained by visible matter alone. Fritz Zwicky’s during the 1930s demonstrated that galaxies must have contained additional "hidden" mass (what we now call DM) otherwise the galaxies themselves would have dissolved.
Then, the seminal observations of galactic rotation curves by Rubin and collaborators revealed nearly flat circular-velocity profiles that extend well beyond the luminous component, in clear tension with expectations based on visible matter alone \cite{Rubin1980,Rubin1986}. These results were subsequently supported by independent evidence from galaxy clusters and gravitational lensing \cite{XXLConsortium:2018dld, DES:2024rfx, DESI:2025zgx, DES:2026fyc}, pointing to the presence of a dominant non-luminous matter component shaping the gravitational potential of galaxies, e.g. see \cite{Salucci:2018hqu}.

A second, conceptually distinct dark component was introduced at the end of the twentieth century with the discovery of the accelerated expansion of the Universe. Observations of distant Type Ia supernovae showed that cosmic expansion is undergoing late-time acceleration \cite{Riess:1998AJ,Perlmutter:1999ApJ}. Within the standard $\Lambda$CDM cosmological model, dark matter and dark energy together account for most of the cosmic energy budget, despite their unknown origins.

Recently, \cite{Kou:2025yfr} emphasized the possibility of explaining the DESI observational measurements [ref.] by considering a component that can unify Cold Dark Matter (CDM) and Dark Energy (DE) into a single entity, capable of explaining both structure formation and accelerated expansion of the Universe.  Two notable properties of these unified models—which we will hereafter refer to as Unified Dark Matter (UDM) models—are: (i) a very small effective speed of sound, which allows for clustering on large scales, and (ii) no violation of the Null Energy Condition (NEC).

The study of UDM models is not recent: we can trace back to the early 2000s literature, where several types of scalar field Lagrangians were considered. However, the Lagrangian that allows a very small speed of sound is that of K-essence; for references, see the review by \cite{Bertacca:2010UDMreview}. Let us briefly review the principal models studied in the literature. At an early stage, unified DM and DE models were formulated at the level of effective fluids. A prominent example is the Chaplygin gas and its generalizations, in which a single fluid interpolates between dust-like behavior at early times and a negative-pressure component at late times. Although these models successfully reproduce a unified dark sector at the background level, they encounter severe difficulties when perturbations and structure formation are considered. In particular, the presence of a non-negligible effective sound speed leads to oscillations or instabilities in the matter power spectrum and prevents the formation of realistic gravitationally bound structures, including galactic haloes \cite{Sandvik:2004ex}. As a result, Chaplygin-type models with a canonical kinetic structure provide an instructive proof of principle for dark-sector unification but fail to account for observed galactic dynamics without fine-tuning.

These limitations motivated the search for more flexible unified descriptions grounded in a field-theoretic framework.
UDM models pursue this idea by describing both dark matter–like clustering and dark energy–like acceleration in terms of a single effective component, typically realised as a scalar field with non-canonical kinetic terms, e.g. K-Essence models.
A consistent realisation of this programme was developed in a sequence of works.
For example, in \cite{BERTACCA_2007}, it was shown that scalar-field dynamics can reproduce a $\Lambda$CDM-like background evolution while remaining compatible with linear perturbation theory.
The impact of the effective sound speed on cosmological observables, in particular the Integrated Sachs-Wolfe effect, was analysed in detail in \cite{Bertacca:2007JCAP_ISW}, leading to stringent viability conditions for UDM scenarios.
Furthermore, in \cite{Bertacca_2008}, the authors generalized the reconstruction of UDM models using a K-essence Lagrangian with a non-trivial, time-dependent equation of state. This allowed for a remarkable feature: while the total equation of state parameter remains greater than minus one (thus preserving the NEC), the effective dark energy component can exhibit an equation of state less than minus one, mimicking phantom behaviour. This reconstruction enables both apparent super-accelerated expansion and an extremely small effective sound speed, thereby allowing structure formation on large scales, without any genuine phantom behaviour or associated instabilities. 
Finally, in \cite{Bertacca:2010mt}, the authors generalized the reconstruction of UDM models using a K-essence Lagrangian with a non-trivial, time-dependent equation of state. This allowed for a remarkable feature: while the total equation of state parameter remains greater than minus one (thus preserving the NEC), the effective dark energy component can exhibit an equation of state less than minus one, mimicking phantom behaviour. This reconstruction enables both apparent super-accelerated expansion and an extremely small effective sound speed, thereby allowing structure formation on large scales, without any genuine phantom behaviour or associated instabilities (see also \cite{Bruni:2012sn, Frion:2023xwq}).

Instead, the extension of the UDM framework to strongly non-linear and virialized regimes was addressed in \cite{Bertacca-Halo_UDM}, where static, spherically symmetric halo configurations were explicitly constructed within scalar-field UDM models.
Inspired by the work of \cite{Armendariz_Picon_2005} and \cite{Bharadwaj:2003iw}, who studied spherically symmetric configurations of K-essence Lagrangians, \cite{Bertacca-Halo_UDM} analysed the rotation velocities of galaxies and found that flat rotation curves at large halo radii are achievable in these models. Crucially, such asymptotic behaviour requires a non-negligible pressure component relative to the energy density. This analysis was further developed in \cite{Bertacca-Halo_UDM}, where various Lagrangian forms and density profiles were systematically explored, with particular attention to reconstructing realistic density profiles consistent with both nearly flat galactic rotation curves and a cosmological constant.
In this case \cite{Bertacca-Halo_UDM} has glimpsed the way to establish which UDMs can have a valid connection between cosmology and phenomenology on a galactic scale (see also the review \cite{Bertacca:2010UDMreview}).
Finally, \cite{Bertacca-Halo_UDM} suggested the possibility of constructing an equivalence class of models that can have the same rotation speed as galaxies but a different density and pressure profile than the CDM. This allows us to reconstruct a non-trivial density and pressure profile for the UDM models.

Subsequently, \cite{Diez-Tejedor:2013sza}, 
starting from Derrick's theorem, they posed a no-go theorem for models with scalar field Lagrangians in static and symmetric configurations or in spherical symmetry.  They demonstrated that such configurations can exhibit negative energy densities—a conclusion already reached in \cite{Bertacca-Halo_UDM}. This potentially poses a significant challenge for UDM models in static configurations. In particular, it was understood that the region most susceptible to negative energy density could be the galactic bulge, where the rotation velocity increases as a function of radius from the galactic center.

However, as we will see in this paper, the crucial quantity to examine is not the energy density itself but rather the NEC, which must be satisfied. We will show, from two complementary perspectives—namely, following \cite{Babichev:2007dw}  and/or working directly within general relativity—when the inertial mass is assumed to be positive (or zero), the necessary and sufficient condition for stable static spherically symmetric configurations is the satisfaction of the NEC. Importantly, this remains valid even in cases where the energy density is negative.

Motivated by these considerations, in this paper we directly consider a phenomenological framework to investigate galactic halos within the UDM paradigm. Precisely, we consider a phenomenological description of the circular velocity profile, designed to reproduce the qualitative behaviour observed across the main galactic regimes: an inner region characterised by a rising velocity, an approximately flat plateau in the intermediate region, and an outer halo region. This phenomenological rotation curve is adopted as a kinematical input and used to reconstruct the effective energy density, pressure components required to support a static halo within a given UDM equivalence class.
In this way, the phenomenological success of standard cold dark matter halo models is retained, while their underlying description is consistently embedded into a relativistic UDM framework.
In the second part of our work, a second method was used to reconstruct a possible pressure and energy density profile of UDM models. Starting directly from the various most commonly used cold dark matter density profiles in the literature, we were also able to reconstruct a UDM profile that describes the same trend as the rotation velocity profile of galaxies.
All these constructions allow us (i) to distinguish which properties are fixed by kinematics and which remain model-dependent within the UDM class, and to systematically explore different halo realizations within a unified description of the dark sector; (ii) even more importantly, to address the stability of UDM models by highlighting that both reconstruction methods require satisfaction of the NEC.

The paper is organised as follows.
In Sec.~\ref{sec:UDMframework} we review the scalar-field UDM setup.
In Sec.~\ref{sec:rotationcurve} we discuss the weak-field connection between circular velocity and the gravitational field and introduce the phenomenological rotation-curve description adopted in this work and, finally we discuss the equivalence-class structure underlying different realisations.
In Sec.~\ref{Phenomenological_analysis-SC}, using the equivalence map between 
CDM and UDM models and starting from a generic circular velocity profile, we 
reconstruct the pressure and energy density of a UDM model sharing the same 
rotation curve.
In Sec.~\ref{sec:profiles} we apply this method to several commonly used halo profiles and analyse the resulting galactic dynamics.
Finally, Sec.~\ref{sec:conclusions} summarises our results and outlines possible extensions.

Finally, throughout the paper, the following conventions will be adopted: $c^2=1$, $8\pi G=1$, and signature $(-,+,+,+)$; Greek indices run over space-time dimensions, while Latin indices label spatial coordinates.

\section{Preliminary setup of UDM models} \label{sec:UDMframework}

The UDM models can provide a simple interpretation of the nature of the 
dark components (i.e., DM and DE) of our Universe at very large scales 
(after the recombination epoch) \cite{Bertacca_2008, Bertacca:2010mt} (see also \cite{Bertacca:2010UDMreview} and all references therein). At the same time, 
at very small scales, UDM models should be able to describe the static 
solutions of Einstein's field equations, e.g.\ see 
\cite{Armendariz_Picon_2005, faber2006galactic, Matos_2000, Hern_ndez_2004, Herrera_2001, Nucamendi_2001, Mak_2003, Diez2006, Bertacca-Halo_UDM, Diez-Tejedor:2013sza}.
Several UDM models are defined starting from a Lagrangian that describes 
dark matter halos in terms of bosonic scalar fields with non-canonical 
kinetic terms; e.g.\ see \cite{Armendariz_Picon_2005, Bertacca_2008, Bertacca-Halo_UDM, Diez-Tejedor:2013sza}.\\

For these specific models, the following UDM action has been considered.
\begin{equation}\label{Eq:action}
    S= S_G+S_\phi+ S_b=\int \ud^4x \sqrt{-g}\left[\frac{R}{2} + 
    \mathcal{L}(\phi, X)\right] + S_b,
\end{equation}
where $S_b$ describes the baryonic matter.
As we can see from Eq.~(\ref{Eq:action}), for simplicity, we assume a 
Lagrangian $\mathcal{L}$ of a scalar field minimally coupled to gravity.
Here, the term $X$ corresponds to the standard kinetic term
\begin{equation}
    X=-\frac{1}{2}g^{\mu\nu}\nabla_\mu \phi \nabla_\nu \phi.
\end{equation}
The stress-energy-momentum tensor of the scalar field can be derived 
from the following expression:
\begin{equation}\label{enmom1}
    T^{\phi}_{\mu\nu}= -\frac{2}{\sqrt{-g}}\frac{\delta S_\phi}
    {\delta g^{\mu\nu}}= \frac{\partial\mathcal{L}(\phi, X)}{\partial X}
    \nabla_\mu \phi \nabla_\nu \phi+\mathcal{L}(\phi, X)g_{\mu\nu}\,.
\end{equation}
Following \cite{Babichev:2007dw}, in this paper, we adopt the Null Energy 
Condition (NEC); i.e., if $n^\mu$ is a null vector, then we have 
$T^{\phi}_{\mu\nu} n^\mu n^\nu \ge 0$, and consequently, this implies that 
\begin{equation}
\frac{\partial\mathcal{L}(\phi, X)}{\partial X} \ge 0\;.
\end{equation}
As pointed out, for example, in \cite{Arefeva:2006ido}, the violation of 
this NEC condition would imply the intrinsic instability of the system; 
see also \cite{Babichev:2007dw}. (We will return to this point in 
Sec.~\ref{consist}.)
Of course, this condition is weaker than the Weak Energy Condition (WEC). 
The WEC guarantees that the energy density (of our scalar field) measured 
along a time-like orbit $t^\mu$ should always be non-negative, i.e.\ 
$\rho_t=T^{\phi}_{\mu\nu} t^\mu t^\nu \ge 0$. In this way, we are able 
to ignore the no-go theorem highlighted in \cite{Diez-Tejedor:2013sza}; 
see also \cite{Diez2006}. 
In a cosmological context, even for the nearly homogeneous Universe 
(e.g., at very large scales), we assume $\nabla_\mu \phi$ is time-like. 
Then we have $X>0$, and the scalar field action describes a perfect fluid 
with (e.g. see  \cite{Mukhanov2007, Bertacca:2010UDMreview})
\begin{equation}
    T^\phi_{\mu\nu}= (\rho_{\rm cosm}+p_{\rm cosm})u_\mu u_\nu+ 
    p_{\rm cosm} g_{\mu\nu}
\end{equation}
where the pressure equals the Lagrangian:
\begin{equation}
p_{\rm cosm}(\phi, X)= \mathcal{L}(\phi, X)
\end{equation}
and energy density
\begin{equation}
\rho_{\rm cosm} (\phi, X)= 2X\frac{\partial \mathcal{L}(\phi, X)}
{\partial X}-\mathcal{L}(\phi, X)\;.
\end{equation}
In this case, $u_\mu=\nabla_\mu \phi/\sqrt{2X}$ is the ``energy frame'' 
four-velocity \cite{Bertacca:2010mt}.
In the opposite limit, namely in a static configuration on smaller scales, 
since we consider a fluid described by a scalar field, there is no 
vorticity and, consequently, we require zero heat flux \cite{EllisBook2012}. 
Following the work of \cite{Herrera:2020wth, Herrera:2022ulc}, the generic 
anisotropic energy-momentum tensor in a static configuration turns out to 
be (e.g., see also \cite{Herrera:1984kfa} and references therein 
\cite{Mak_2003})
\begin{equation}\label{generalT}
    T_{\mu\nu}=(\rho+p_\perp)t_\mu t_\nu+ p_\perp g_{\mu\nu}+
    (p_\parallel-p_\perp)s_\mu s_\nu.
\end{equation}
where $s^\mu$ is a space-like unit vector field, $\rho$ is the energy 
density, $p_\parallel$ is the radial pressure, and $p_\perp$ is the 
orthogonal pressure, i.e., in the direction orthogonal to $s^\mu$ and 
$t^\mu$. (Here $t^\mu$ is a time-like unit vector field of a generic 
static observer.) 
Assuming a static, spherically symmetric configuration, the scalar field 
depends only on spatial coordinates; i.e., we have $X<0$. Then we can 
directly identify the space-like vector as $s_\mu= \nabla_\mu \phi/
\sqrt{-2X}$ and the four-velocity of the observer at rest frame as 
$t^\mu=\delta^\mu_0/\sqrt{-g_{00}}$, where $g_{00}$ is the $00$ 
component of the metric.
In this case, by comparing the stress-energy-momentum tensor of the 
scalar field with the one in Eq.~(\ref{generalT}), we conclude that 
\begin{equation}
\begin{split}\label{relation} 
      \rho= -p_\perp = -\mathcal{L}, \qquad p_\parallel= \mathcal{L}-
      2X\frac{\partial \mathcal{L}}{\partial X}.\\
\end{split}
\end{equation}   
The stress-energy-momentum tensor simplifies as follows
\begin{equation}\label{enmom2}
        T^\phi_{\mu\nu} =(p_{\parallel}+\rho)s_\mu s_\nu + 
        p_\perp g_{\mu\nu}.
\end{equation}

In the following sections, we consider a general metric of a spherically 
symmetric static spacetime. Following the analysis of \cite{Bertacca-Halo_UDM}, 
we study in detail whether dark matter halos can be described in terms of 
bosonic scalar fields with non-canonical kinetic terms; see also 
\cite{Bertacca_2008, Armendariz_Picon_2005, Diez-Tejedor:2013sza}.
At this point, it is possible to introduce one of the most common 
expressions of a general static, spherically symmetric line element.
\begin{equation}\label{metric}
    \ud s^2= -e^{2\alpha(r)} \ud t^2+ e^{2\beta(r)}\ud r^2 + 
    r^2(d\theta^2+\sin{\theta}^2 \ud \psi^2)
\end{equation}
and the $(tt)$, $(rr)$ and $(\theta\theta)$ components of the Einstein 
equations are
\begin{equation}\label{Albert1}
\begin{split}
    &\quad \frac{1}{r^2}\left[ e^{-2\beta(r)}-2re^{-2\beta(r)}\beta'-1\right]= -\rho,\\
    &\quad \frac{1}{r^2}\left[ e^{-2\beta(r)}+2re^{-2\beta(r)}\alpha'-1\right]= p_\parallel,\\
    &\quad e^{-2\beta(r)}\left[ \alpha''+ (\alpha')^2-\alpha'\beta'\right]+ \frac{e^{-2\beta(r)}}{r}\left[ \alpha'-\beta'\right]= p_\perp=-\rho.\\
    \end{split}
\end{equation}

Now, for a generic scalar field Lagrangian, given the covariant 
conservation of the stress-energy tensor $\nabla_\mu T^{\mu\nu}=0$, 
the equation of motion for a scalar field in a generic metric is
\begin{equation}\label{scalar}
\nabla_\mu\left[\frac{\partial\mathcal{L}(\phi, X)}{\partial X}
\left(-\partial^\mu\phi\right)\right]-\frac{\partial\mathcal{L}(\phi, X)}
{\partial \phi}=0\;.
\end{equation}
Starting from this equation of motion, we can rewrite it as the 
hydrostatic equilibrium equation, i.e.,
\begin{equation}\label{cont}
p_\parallel'= -(p_\parallel+\rho)\left[\frac{2}{r}+\alpha'\right],
\end{equation}
where a prime denotes the derivative with respect to $r$.
Note that this reduces to the Tolman-Oppenheimer-Volkoff (TOV) equation 
when $\rho=-p_\perp$, and for hydrostatic equilibrium, we require 
$p_\parallel \ge 0$.
Using the spherically symmetric metric defined in Eq.~(\ref{metric}), 
the kinetic term can be explicitly written in terms of the metric 
functions and the scalar field as follows
\begin{equation}\label{X}
    X(r) = -\frac{1}{2}\left(\dfrac{d\phi}{dr}\right)^2 e^{-2\beta}, 
    \qquad \phi(r, r_{\rm min}) = \int^r_{r_{\rm min}} \sqrt{-2X(\tilde r)} 
    e^{2\beta(\tilde r)} \ud \tilde r\,
\end{equation} 
where we can set $r_{\rm min} \ge 0$.

An alternative metric parametrisation is useful to highlight the physical 
meaning of the metric functions by introducing the quantity $m(r)$, which 
represents the Hernandez-Misner/Misner-Sharp quasi-local mass 
\cite{Misner_1964, Hernandez1966}. It corresponds to the effective 
gravitational mass within radius $r$ defined in \cite{hawking1975large}.
This metric is given by the following line element
\begin{equation}
    ds^2 = -\frac{e^{2R(r)}}{(r/r_0)^4}dt^2 + \frac{1}{1 - 
    \frac{m(r)}{4\pi r}}dr^2 + r^2(d\theta^2 + \sin^2{\theta} d\psi^2),
\end{equation}
where $r_0$ is a suitable length-scale, $R = \ln[(r/r_0)^2 \exp{\alpha(r)}]$ 
and $e^{-2\beta(r)} = 1 - m(r)/4\pi r$.\\
From the Einstein equations, it follows that
\begin{equation}\label{mass}
    m' = 4\pi \rho r^2, \qquad \qquad m(r) - m(r_0) = 4\pi \int_{r_0}^r 
    \tilde{r}^2 \rho(\tilde{r}) d\tilde{r},
\end{equation}
The following relations between the two metrics hold:
\begin{equation}\label{der_alpha}
  \beta(r)= \frac{1}{2}\ln{\frac{1}{ 1 - m(r)/4\pi r}}, \qquad\qquad  
  \alpha' = R' - \frac{2}{r} = \frac{m/8\pi + p_\parallel r^3/2}
  {r^2\left[1 - m/4\pi r\right]}.
\end{equation}
Then, in terms of the function $R$, the covariant conservation of the 
stress-energy tensor becomes:
\begin{equation}\label{ContEq}
    \frac{dp_\parallel}{dR} = -(p_\parallel + \rho).
\end{equation}

\subsection{Consistency conditions}\label{consist}

To ensure the physical viability of our reconstructed UDM model, we have 
to impose several fundamental consistency conditions. These requirements 
ensure that the model possesses the qualities and properties necessary to 
correctly describe DM halos, while avoiding pathological behavior.

\begin{enumerate}
\item As we pointed out before, the NEC represents a minimal requirement 
for stability; e.g., see \cite{Babichev:2007dw}. For any null vector 
$n^\mu$ ($n_\mu n^\mu = 0$), the NEC requires:
\begin{equation}
T_{\mu\nu}n^\mu n^\nu \geq 0,
\end{equation}
which for our scalar field Lagrangian $\mathcal{L}(\phi, X)$ translates to:
\begin{equation}
X\frac{\partial\mathcal{L}(\phi, X)}{\partial X} \ge 0.
\end{equation}
This condition is deeply connected to the absence of ghosts and gradient 
instabilities: violating the NEC would lead to an unbounded Hamiltonian 
from below, resulting in vacuum instabilities that would render the theory 
unphysical. Both aspects are thus two sides of the same coin, and together 
they ensure the physical viability of the reconstructed UDM model, 
particularly in static, spherically symmetric configurations where the 
kinetic term $X$ becomes negative.

\item The sound speed of scalar perturbations must satisfy:
\begin{equation}\label{sound}
c_s^2 = \frac{\partial\mathcal{L}/\partial X}{\partial\mathcal{L}/\partial X 
+ 2X\partial^2\mathcal{L}/\partial X^2} > 0.
\end{equation}
Following \cite{Babichev:2007dw}, this condition is essential to avoid 
catastrophic instabilities in non-linear regimes, particularly in 
high-density regions.

\item For the theory to possess a well-posed initial value formulation 
with a globally hyperbolic spacetime structure (where Cauchy surfaces 
exist and causality is preserved \cite{Wald1984}), additional constraints 
emerge:
\begin{equation}\label{con1}
X\frac{\partial\mathcal{L}}{\partial X} > 0, \qquad 
\frac{\partial\mathcal{L}}{\partial X} + 2X\frac{\partial^2\mathcal{L}}
{\partial X^2} > 0.
\end{equation}
These conditions ensure the absence of ghosts and gradient instabilities 
in the low-energy spectrum of the theory \cite{Diez-Tejedor:2013sza}, and 
are intimately related to the NEC discussed in point 1: both sets of 
conditions reflect the requirement that the theory remains physically 
consistent, free from pathological behaviour, and causally well-posed. 
As shown in \cite{Babichev:2007dw}, imposing the first and second 
conditions automatically satisfies the third. Furthermore, these 
constraints guarantee the hyperbolic nature of the equation of motion, 
ensuring causal propagation and keeping the Cauchy problem well-posed.

\end{enumerate}

These consistency conditions are particularly crucial in our reconstruction, 
especially within the region of the Galactic Bulge where ${(v_c^2)}' > 0$, 
because they prevent the emergence of pathological configurations that might 
otherwise occur in the central density profiles.
In particular, they allow us to extend our UDM model reconstruction to the 
entire halo, remaining free of ghosts, tachyons, and gradient instabilities, 
while maintaining causal energy propagation. This provides essential 
theoretical priors that constrain the functional form of $\mathcal{L}(\phi, X)$ 
during the reconstruction process from observational data, in particular when 
dealing with the core-cusp problem in galactic centres (e.g., see 
\cite{de_Blok_2009} and references therein).
Finally, we note that in \cite{Sawicki:2024ryt} the authors generalised the 
above discussion to any scalar-tensor theory, deriving the conditions for a 
given model to admit a Cauchy surface on which a well-posed initial value 
problem can be formulated.

\subsection{NEC through the positive condition of the inertial mass}

Before going into the main topic of this paper, let us briefly discuss 
the importance of satisfying the NEC.
As pointed out in~\cite{Rosen_1992, EllisBook2012}, the inertial mass 
density of matter is given by the sum of $\rho + p$, where
\begin{equation}
p=\frac{1}{3} \left(2 p_\perp +p_\| \right)\;.
\end{equation}
In our case, $p_\perp =-\rho$, which is equivalent to 
$\rho + p_\| \ge 0$.
In particular,~\cite{EllisBook2012} pointed out that this quantity 
must be positive because it is related to any form of internal energy 
of the fluid (e.g., heat or chemical energy). They further conclude 
that this contributes to a correct effective inertial mass, ensuring 
that matter tends to move in the direction of an applied pressure 
gradient, rather than in the opposite direction.

In conclusions, the crucial point is to satisfy this relation in order to 
have a stable solution. This is our starting point, and in the next 
sections we will show how the NEC can be preserved, both for a generic 
velocity curve and starting from a generic CDM profile.

\section{Circular velocities and Rotation curve}\label{sec:rotationcurve}

In Newtonian gravity, the circular velocity \( v_c \) of a test 
particle in a spherically symmetric potential is given by the 
simple relation:
\begin{equation} \label{Eq:rotation_curves-Newtonian_def}
    v_c^2(r) = \frac{ M^{\rm CDM}(r)}{8 \pi r},
\end{equation}
where \( M^{\rm CDM}(r) \) is the total halo mass enclosed within 
radius \( r \). For spiral galaxies, outside the galactic bulge, 
most stars are concentrated in a thin disc. The rotation velocity 
of the stars traces the Cold Dark Matter (CDM) halo through their 
orbital motion within the galactic plane (\( \theta = \pi/2 \)). 
Observations show that the rotation curves remain nearly flat 
(\( v_c \approx \text{constant} \)) at large radii, where visible 
matter becomes negligible. This has provided one of the most 
convincing pieces of evidence for the existence of dark matter 
halos; see~\cite{Rubin1980}.

In the UDM framework proposed by~\cite{BERTACCA_2007}, a single 
exotic fluid accounts for both Dark Matter (DM) and, at 
cosmological scales, Dark Energy (DE). In UDM models, as we have 
already emphasised previously, we have not only a non-negligible 
energy density but also a non-negligible radial pressure. 
For this reason, the general relativistic expression for the 
circular velocity in a static and spherically symmetric spacetime, 
obtained from the geodesic equation of an observer moving 
circularly with respect to the centre of the halo (which we assume 
to coincide with that of the spiral galaxy), contains additional 
terms due, precisely, to the pressure of the fluid itself. 
Following~\cite{Bertacca-Halo_UDM}, we have
\begin{equation}
    v_c^2(r) = \frac{p_\parallel r^2/2 + m/(8\pi r)}
    {1 - \left[p_\parallel r^2/2 + 3m/(8\pi r)\right]},
\end{equation}
where \( m = m_{\text{UDM}} + m_{\rm B} \) represents the total mass 
within radius \( r \), including both the UDM fluid and baryonic 
components. The baryonic mass \( m_{\rm B}  \) is typically concentrated 
within a central scale radius \( r_{\rm B} \) (e.g., the bulge region).

For non-relativistic orbital velocities (\( v_c \ll 1 \)) and in 
the weak-field limit (\( m/(8\pi r) \ll 1 \) and 
\( p_\parallel r^2 \ll 1 \)), this reduces to the modified 
expression:
\begin{equation}\label{va}
    v_c^2(r) \approx \frac{m}{8\pi r} + \frac{p_\parallel r^2}{2}.
\end{equation}
Notably, this maintains the Newtonian form plus a pressure-dependent 
correction, reflecting the non-trivial equation of state of the UDM 
fluid. A crucial insight from~\cite{BERTACCA_2007} is that the 
rotation curve in Eq.~\eqref{va} remains invariant under the 
transformation:
\begin{equation}\label{goodtrans}
    \begin{cases}
        \rho \longrightarrow \tilde{\rho} = \rho + \sigma(r), \\
        p_\parallel \longrightarrow \tilde{p}_\parallel = 
        p_\parallel + \lambda(r),
    \end{cases}
\end{equation}
provided the functions \( \sigma(r) \) and \( \lambda(r) \) 
satisfy:
\begin{equation}\label{diff_goodtrans}
    3\lambda(r) + r \lambda'(r) = -\sigma(r)\;.
\end{equation}
This relation is useful for our purposes because it allows us to 
consider different static solutions with appropriate energy 
densities $\rho$ and pressures $p_\|$, which give rise to the 
same rotation curve (i.e., the same $v_c$), always in the case 
of weak fields. Here, an important comment is in order.
 
Obviously, assuming a UDM reconstruction that only works in the weak-field 
regime, we cannot study this model with our approach for $r \to 0$. 
For simplicity, we therefore impose that $\rho$ and $p_\|$ are negligible 
in the central region of the halo, and assume the existence of a central 
compact object which can be treated as a separate $\delta$-function-like 
point mass of mass $m^\ast$ (e.g., a supermassive black hole), dynamically 
distinct from the halo UDM component.

By combining Eqs.~(\ref{ContEq}) and~(\ref{diff_goodtrans}), we obtain a 
useful differential equation that allows us to reconstruct our UDM models. 
To this end, we first need to determine the $p^{\rm UDM}_{\|}$ and 
$\rho^{\rm UDM}$ components. In particular, we are able to obtain our UDM 
model either starting from the observed $v_c$, or directly from the CDM 
halo profile.

Assuming that $v_c^2\simeq r \alpha' \ll 1$, we have
\[ R'={1+r \alpha' \over r} \simeq {2 \over r}\]
and setting $\lambda=p^{\rm UDM}_{\, \|}$ and $\sigma=\rho^{\rm UDM}-\rho^{\rm CDM}$, 
Eqs.~(\ref{ContEq}) and~(\ref{diff_goodtrans}) become 
\begin{equation} \label{Eq:dif_eq1}
  { \ud p^{\rm UDM}_\| \over \ud r} = -{2\over r}(p^{\rm UDM}_\| + \rho^{\rm UDM})\;,
\end{equation}
and
\begin{equation} \label{Eq:dif_eq2}
  3p^{\rm UDM}_\|+ r \frac{\ud  p^{\rm UDM}_\|}{\ud r} =
  -\left(\rho^{\rm UDM}-\rho^{\rm CDM}\right)\;.
\end{equation}
Here, the UDM and CDM indices refer to the pressure and density of the 
UDM and CDM models, respectively.
Then, combining these differential equations --- e.g., by eliminating 
$\rho^{\rm UDM}$ --- we find
\begin{equation}
    { \ud \over \ud r} \left(r^4 p^{\rm UDM}_\|\right)=2  r^3\rho^{\rm CDM} \;,
\end{equation}
and the following solution
\begin{equation}
  p^{\rm UDM}_\|(r) ={2 \over r^4} \int^r \tilde r^3 \rho^{\rm CDM} 
  (\tilde r)\, \ud \tilde r + {K \over r^4} \,.
\end{equation}
Using either of Eqs.~(\ref{Eq:dif_eq1}) and~(\ref{Eq:dif_eq2}), we find
\begin{equation}
   \rho^{\rm UDM}(r) =  - \rho^{\rm CDM} + {2 \over r^4} \int^r \tilde r^3 
   \rho^{\rm CDM} (\tilde r)\, \ud \tilde r + {K \over r^4}= 
   p^{\rm UDM}_\|(r)- \rho^{\rm CDM}\;.
\end{equation}
Finally, in order to satisfy the NEC we have to impose
\begin{equation}
   p^{\rm UDM}_\|(r)\ge {\rho^{\rm CDM} \over 2}\,.
\end{equation}
In relation to the additive term $K/r^4$ contained in $p^{\rm UDM}_\|$ 
and $\rho^{\rm UDM}$, one comment is in order.
If we consider a definite integral, then we can rewrite $\rho^{\rm UDM}(r)$ 
and $p^{\rm UDM}_\|$ in the following way 
\begin{equation}\label{eq:p_par-int_def-1}
  p^{\rm UDM}_\|(r) ={2 \over r^4} \int_{r_{\rm min}}^r \tilde r^3 
  \rho^{\rm CDM} (\tilde r)\, \ud \tilde r + 
  {p^{\rm UDM}_\|(r_{\rm min}) \over r^4} \,.
\end{equation}
Using either of Eqs.~(\ref{Eq:dif_eq1}) and~(\ref{Eq:dif_eq2}), we find
\begin{equation}\label{eq:rho-int_def-1}
   \rho^{\rm UDM}(r) =  - \rho^{\rm CDM} + {2 \over r^4} \int_{r_{\rm min}}^r 
   \tilde r^3 \rho^{\rm CDM} (\tilde r)\, \ud \tilde r + 
   {p^{\rm UDM}_\|(r_{\rm min}) \over r^4}\;.
\end{equation}
Here the physical meaning of $r_{\rm min}$ can be related to the method 
used to reconstruct the UDM model. For example, if we consider the 
prescription using the observed rotation curves, we can set $r_{\rm min}$ 
as the minimum radius that has been measured. Obviously, if we consider 
the reconstruction using the CDM density contrast, $r_{\rm min}$ could 
also be related to the radius at which the weak-field approximation can 
no longer be applied.

In the following section, we apply the approach outlined in the above comment
and, in particular, we study the reconstruction of the UDM model using $v_c$ directly.

\section{Phenomenological analysis of galactic rotation curves}
\label{Phenomenological_analysis-SC}

he rotation curve observed in a generic spiral galaxy, $v_c(r)$, presents 
two main regimes that reflect the underlying mass distribution: the bulge 
region, near the centre, and the plateau/disk region, extending up to about 
$10$--$20$ kpc depending on the size of the galaxy. Since the different 
behaviours of $v_c(r)$ determine different matter distributions in these 
regions, valid UDM models may exhibit entirely different regimes in the 
bulge region compared to the plateau at larger $r$.

\begin{figure}[h!]
\centering
    {\centering\Large\textbf{rotational velocities}\par\medskip}
        \includegraphics[width=\linewidth]{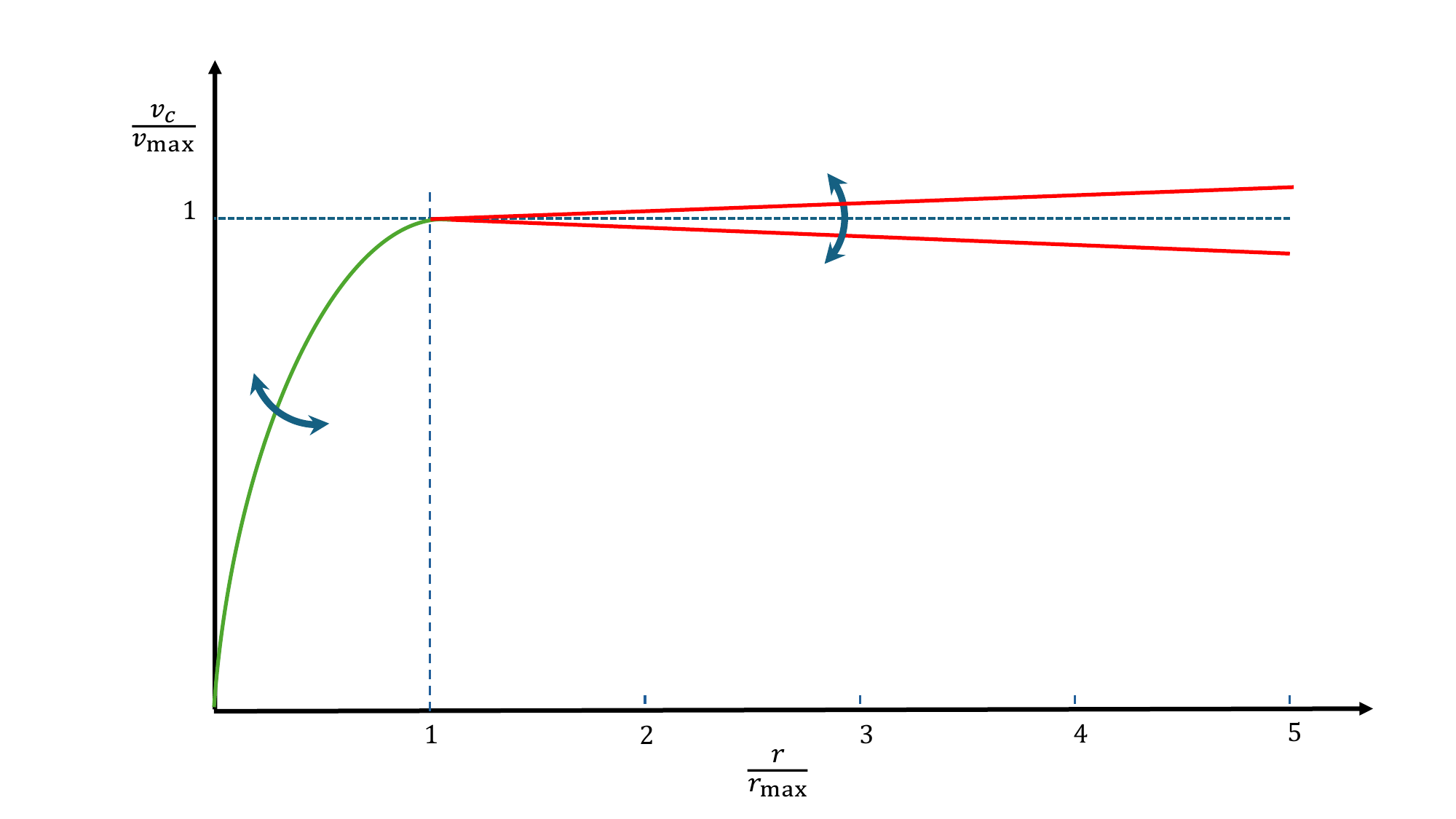}
    \caption{In this Figure we show a suitable parametrisation of the rotation curve of a spiral galaxy.
    }
    \label{fig:rotational_velocities}
\end{figure}

As a visual representation of a typical profile of a galactic rotation curve, let us consider a generic prototype, as shown in Fig. \ref{fig:rotational_velocities}. 
Let us consider a rotation curve of such a dark matter halo in observed 
units $[r/r_{\rm max}, v_c/v_{\rm max}]$, where the point $[1, 1]$ 
separates, at first approximation, the bulge and the disc regions. 
Here, $v_{\rm max}$ is the maximum velocity and $r_{\rm max}$ is the 
radius of the turnover velocity. Splitting it into these regions, we can 
conveniently parametrise $v_c/v_{\rm max}$ in the following way
\begin{itemize}
    \item{\bf Bulge region}:
    \begin{equation} \label{eq:bulge_param}
        {v_c\over v_{\rm max}} =1- \left( 1- {r\over r_{\rm max}}\right)^\alpha\,
\end{equation}
    where $\alpha>1$ and $r \le r_{\rm max}$;
    \item{\bf Disk region}:
     \begin{equation} \label{eq:disc_param}
         {v_c\over v_{\rm max}}=1 - \beta \left( {r \over r_{\rm max}}-1\right)^\gamma\,
    \end{equation}
    where $|\beta|\ll1$, $\gamma\ge1$ and $r \ge r_{\rm max}$.
\end{itemize}

Now, considering Eq.(\ref{Eq:rotation_curves-Newtonian_def}), we can write the CDM energy density as
\begin{equation}
    \rho^{\rm CDM} ={1 \over r^3}\left[{\ud \over \ud r}\left(r^2 v_c^2 \right) - r v_c^2 \right]\;,
\end{equation}
and using  Eqs. (\ref{eq:p_par-int_def-1}), we can write $p^{\rm UDM}_\|(r)$  in the following way
\begin{equation}\label{eq:p_par-int_def-2}
  p^{\rm UDM}_\|(r) ={2 \over r^4} \left[r^2 v_c^2(r) -  \int_{r_{\rm min}}^r \tilde r v_c^2(\tilde r)\, \ud \tilde r \right] + %
  {\left[ r_{\rm min}^4  p^{\rm UDM}_\|(r_{\rm min}) -2  r_{\rm min}^2 v_c^2 (r_{\rm min}) \right]  \over r^4} \,
\end{equation}
and Eq.(\ref{eq:rho-int_def-1}) turns our
\begin{equation}\label{eq:rho-int_def-2}
  \rho^{\rm UDM}(r) =    { 1 \over r^4} \left[r^2 v_c^2(r) - 2 \int_{r_{\rm min}}^r \tilde r v_c^2(\tilde r)\, \ud \tilde r \right] -{ 1 \over r}{\ud v_c^2 (r)\over \ud r} + {\left[r_{\rm min}^4 p^{\rm UDM}_\|(r_{\rm min}) -2  r_{\rm min}^2 v_c^2 (r_{\rm min}) \right] \over r^4}\;.
   \end{equation}
   In this specific case, the NEC becomes
   \begin{equation}\label{eq:NEC-int_def-2}
   { 1 \over r^4} \left[3r^2 v_c^2(r) - 4 \int_{r_{\rm min}}^r \tilde r v_c^2(\tilde r)\, \ud \tilde r \right] -{ 1 \over r}{\ud v_c^2 (r)\over \ud r} + 2 {\left[r_{\rm min}^4 p^{\rm UDM}_\|(r_{\rm min}) -2  r_{\rm min}^2 v_c^2 (r_{\rm min}) \right] \over r^4}\ge 0 \;.
   \end{equation}

Starting from the parametrization of the bulge region suggested in 
Eq.~(\ref{eq:bulge_param}), let us consider the case where 
$r_{\rm min} \le r \ll r_{\rm max}$. Then, in this limit, we have
\begin{equation}
v_c^2(r\ll r_{\rm max})\simeq \alpha^2 v_{\rm max}^2 \left({r\over r_{\rm max}}\right)^2 
\quad \quad {\rm and } \quad\quad \rho^{\rm CDM} (r\ll r_{\rm max}) \simeq 3  
{\alpha^2 v_{\rm max}^2 \over  r_{\rm max}^2}\;.
\end{equation}
Therefore, using Eqs.~(\ref{eq:p_par-int_def-2}) and~(\ref{eq:rho-int_def-2}) 
we find
\begin{equation}\label{eq:p_par-lim_r_small}
   p^{\rm UDM}_\|(r\ll r_{\rm max})\simeq {3 \over 2}  
   {\alpha^2 v_{\rm max}^2 \over  r_{\rm max}^2} \left[1 - \left( {r_{\rm min} 
   \over r } \right)^4 \right] +p^{\rm UDM}_\|(r_{\rm min})\left( {r_{\rm min} 
   \over r } \right)^4 
\end{equation}
and 
\begin{equation}\label{eq:rho-lim_r_small}
   \rho^{\rm UDM}(r\ll r_{\rm max})\simeq - {3 \over 2}  
   {\alpha^2 v_{\rm max}^2 \over  r_{\rm max}^2} \left[1 + \left( {r_{\rm min} 
   \over r } \right)^4 \right] +p^{\rm UDM}_\|(r_{\rm min})\left( {r_{\rm min} 
   \over r } \right)^4\;.
\end{equation}
Finally, from the NEC, see Eq.~(\ref{eq:NEC-int_def-2}), the most 
problematic case occurs at $r=r_{\rm min}$ and we obtain the following 
inequality
\begin{equation}\label{eq:NEC-lim_r_small}
   p^{\rm UDM}_\|(r_{\rm min}) \ge {3 \over 2}  
   {\alpha^2 v_{\rm max}^2 \over r_{\rm max}^2} 
\end{equation}
which highlights the necessary and sufficient condition that these UDM 
models must satisfy near the centre of the Bulge region. Moreover, this 
NEC condition is particularly interesting as it relates the rotation 
velocity gradient to the maximum velocity $v_{\rm max}$ (with the 
corresponding radius $r_{\rm max}$) and the pressure value that the 
model must satisfy near the centre of the halo. At this point, let us 
add three more important comments.
   
First of all, it is worth noting that a possible solution near the centre of 
the halo may be related to the superradiance mechanism, e.g.\ see 
\cite{Brito:2015oca}. 
In this context, the term containing $p^{\rm UDM}_\| (r_{\rm min})$ 
can be interpreted and connected to the term proportional to 
$K/r^4$. In this regime, this contribution can be large enough 
to resolve the NEC problem while, at the same time, not 
affecting the measured value of the rotation velocity at 
$r = r_{\rm min}$. However, a detailed analysis of this 
scenario goes beyond the scope of the present paper and 
will be addressed in a future work.

\begin{figure}[h!]
\centering
    {\centering\Large\ \par\medskip}
        \includegraphics[width=\linewidth]{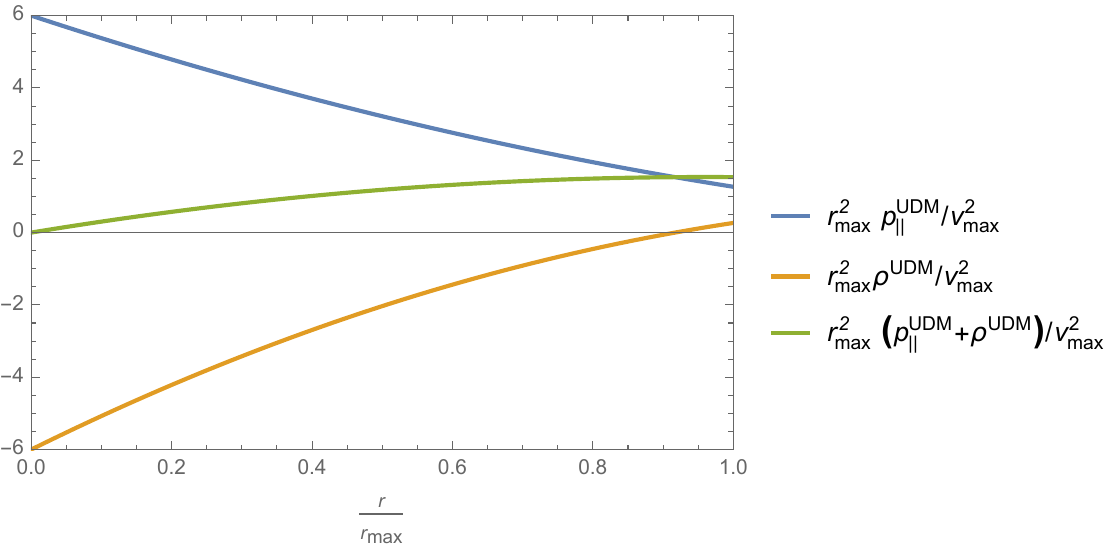}
    \caption{  Pressure and energy density of a UDM model reconstructed from a phenomenological circular velocity profile.  }
    \label{Fig2}
\end{figure}

   Finally, if we consider the case in which we relax $r_{\rm min}=0$, $v_c(0)=0$ and $ p^{\rm UDM}_\|(0)=0$ we find
   \begin{equation}\label{eq:p_par-int_def-rmin=0}
  p^{\rm UDM}_\|(r) ={2 \over r^4} \left[r^2 v_c^2(r) -  \int_{0}^r \tilde r v_c^2(\tilde r)\, \ud \tilde r \right] \,
\end{equation}
\begin{equation}\label{eq:rho-int_def-rmin=0}
  \rho^{\rm UDM}(r) =    { 1 \over r^4} \left[r^2 v_c^2(r) - 2 \int_{0}^r \tilde r v_c^2(\tilde r)\, \ud \tilde r \right] -{ 1 \over r}{\ud v_c^2 (r)\over \ud r} \;,
   \end{equation}
   for the pressure and energy density, and NEC turns out
   \begin{equation}\label{eq:NEC-int_def-rmin=0}
   { 1 \over r^4} \left[3r^2 v_c^2(r) - 4 \int_{0}^r \tilde r v_c^2(\tilde r)\, \ud \tilde r \right] -{ 1 \over r}{\ud v_c^2 (r)\over \ud r} \ge 0\;.
   \end{equation}
Then, e.g., using again the parametrisation defined in 
Eq.~(\ref{eq:bulge_param}), for $r \ll r_{\rm max}$ the NEC is exactly 
zero. This means that in this case we are preserving the model from 
instabilities.
Analytically, pressure and energy density become
\begin{eqnarray}\label{eq:}
  p^{\rm UDM}_\|(r) &=&- {v_{\rm max}^2 r_{\rm max}^2 \over  r^4}\Bigg\{-2 \left(r \over r_{\rm max}\right)^2 \left[\left(1-{r\over r_{\rm max}}\right)^{\alpha }-1\right]^2+\left(r\over r_{\rm max}\right)^2
  \nonumber \\
&&  + {1 \over (\alpha +1) (2 \alpha +1) } \left[- \left( (2 \alpha +1) {r\over r_{\rm max}} +1 \right) \left( 1 - {r\over r_{\rm max}}\right)^{2\alpha +1}+1\right] \nonumber\\
&&  + {4 \over (\alpha +1) ( \alpha +2) } \left[\left( ( \alpha +1) {r\over r_{\rm max}} +1 \right) \left( 1 - {r\over r_{\rm max}}\right)^{\alpha +1}-1 \right]
   \Bigg\}
 \end{eqnarray}
 and
\begin{eqnarray}\label{eq:general_case}
  \rho^{\rm UDM}(r) &=& -{v_{\rm max}^2 r_{\rm max}^2 \over  r^4} 
  \Bigg\{2 \alpha  \left({r\over r_{\rm max}}\right)^3 \left[1-
  \left(1- {r\over r_{\rm max}}\right)^{\alpha }\right] \left(1-{r\over r_{\rm max}}\right)^{\alpha -1} \nonumber\\
  &&-\left({r\over r_{\rm max}}\right)^2
  \left[ \left(1-{r\over r_{\rm max}}\right)^{\alpha }-1\right]^2  +\left({r\over r_{\rm max}}\right)^2\nonumber\\
&& + {1 \over (\alpha +1) (2 \alpha +1) } \left[- \left( (2 \alpha +1) {r\over r_{\rm max}} +1 \right) \left( 1 - {r\over r_{\rm max}}\right)^{2\alpha +1}+1\right] \nonumber\\
&&  + {4 \over (\alpha +1) ( \alpha +2) } \left[\left( ( \alpha +1) {r\over r_{\rm max}} +1 \right) \left( 1 - {r\over r_{\rm max}}\right)^{\alpha +1}-1 \right]\;.
   \end{eqnarray}
Finally, in Figure~\ref{Fig2} we show both quantities and their sum. 
In this case, we can confirm that the NEC is satisfied in the bulge region.
   
    Now, in the junction at $r=r_{\rm max}$, a possible relationship between 
the constants we have introduced in both bulge and disk regions is required. 
In this case, it is necessary to generalise, for example, the above relation 
in the following way
\begin{equation} \label{Eq:general}
     {v_c\over v_{\rm max}} =1- |\beta| {\left[\epsilon_\beta 
     \left(r/r_{\rm max}\right)^\delta +1\right]\over |\beta| +  
     \left(r/r_{\rm max}\right)^\delta } \left| 1- {r\over r_{\rm max}}
     \right|^\alpha\,
\end{equation}
where $\epsilon_\beta ={\rm Sign} \beta$ and $\delta > 1$. Then, if 
$\alpha=\gamma$, Eq.~(\ref{Eq:general}) can reproduce the analysis made 
above at $r \ll r_{\rm max}$ and the disk region for 
$1 < r/r_{\rm max} \lesssim 10$. Furthermore, from now on, for the 
analysis of the disk region, we set directly $\alpha=\gamma$.

    In the disk region, considering only terms up to linear order to $\beta \ll 1$, we find
         \begin{eqnarray}
         p^{\rm UDM}_\|(r) &\simeq& \frac{v_{\max}^2 r_{\rm max}^2}{r^4}\Bigg\{
         \left({ r_{\rm min} \over r_{\rm max}} \right)^2
         + {1 \over (\alpha +1)} \Bigg[ - \frac{\left[1- \left( r_{\rm min}/r_{\rm max} \right) \right]^{2\alpha +1} \left[ \left(2 \alpha +1 \right)\left(r_{\rm min}/r_{\rm max} \right)+1 \right]}{(2\alpha+1)} \nonumber \\
          &&+ 4 \frac{\left[1- \left( r_{\rm min}/r_{\rm max} \right) \right]^{\alpha +1} \left[ \left( \alpha +1 \right)\left(r_{\rm min}/r_{\rm max} \right)+1 \right]}{(\alpha+2)}
         \Bigg]
         +\left(r \over r_{\rm max} \right)^2 -  {4 \beta \over \left(2 +3 \alpha + \alpha^2 \right)}    \times \nonumber \\
        && \quad \times \left[\left(r\over r_{\rm max}\right)-1\right]^{\alpha} \Bigg[
        1 + \alpha \left(r \over r_{\rm max} \right) +\left(1 +\alpha\right)^2 \left(r \over r_{\rm max} \right)^2
        \Bigg] \Bigg\} \nonumber\\
        && +   {\left[ r_{\rm min}^4  p^{\rm UDM}_\|(r_{\rm min}) -2  r_{\rm min}^2 v_c^2 (r_{\rm min}) \right]  \over r^4} +O\left(\beta ^2\right)
    \end{eqnarray}
and 
    \begin{eqnarray}
    \rho^{\rm UDM}(r) &\simeq& 
    \frac{v_{\max}^2 r_{\rm max}^2}{r^4}\Bigg\{ \left({ r_{\rm min} \over r_{\rm max}} \right)^2
    + {1 \over (\alpha +1)} \Bigg[ - \frac{\left[1- \left( r_{\rm min}/r_{\rm max} \right) \right]^{2\alpha +1} \left[ \left(2 \alpha +1 \right)\left(r_{\rm min}/r_{\rm max} \right)+1 \right]}{(2\alpha+1)} \nonumber \\
          &&+ 4 \frac{\left[1- \left( r_{\rm min}/r_{\rm max} \right) \right]^{\alpha +1} \left[ \left( \alpha +1 \right)\left(r_{\rm min}/r_{\rm max} \right)+1 \right]}{(\alpha+2)}
         \Bigg]  +  {2 \beta \over \left(2 +3 \alpha + \alpha^2 \right)}    \times \nonumber\\        
         &&    \times  \left[\left(r\over r_{\rm max}\right)-1\right]^{\alpha} \left[\alpha (\alpha+1)^2 \left(r \over r_{\rm max} \right)^3  + \alpha (\alpha -1)\left(r \over r_{\rm max} \right)^2 +2 (\alpha-1)\left(r \over r_{\rm max} \right) +2 \right]\Bigg\} \nonumber \\
 && +   {\left[ r_{\rm min}^4  p^{\rm UDM}_\|(r_{\rm min}) -2  r_{\rm min}^2 v_c^2 (r_{\rm min}) \right]  \over r^4} +O\left(\beta ^2\right)\;.
    \end{eqnarray}
In this equation, we have split the integral into two parts: in the first, 
from $r_{\rm min}$ to $r_{\rm max}$, we have considered the rotation 
velocity defined in the bulge region, i.e.\ Eq.~(\ref{eq:bulge_param}), 
and for $r \ge r_{\rm max}$ we have used Eq.~(\ref{eq:disc_param}) 
(with $\alpha=\gamma$).
For large $r/r_{\rm max}$, by considering $\alpha \ge 1$ and discarding 
terms of order $O\left((r_{\rm max}/r)^4\right)$, we get the following 
simplified relation
   \begin{eqnarray}
         p^{\rm UDM}_\|(r) &\simeq& {v_{\max}^2 \over  r_{\rm max}^2}\left[\left(r_{\rm max} \over r \right)^2  -4 \beta { \left(1 +\alpha\right)  \over \left(2 + \alpha  \right)}   \left(r\over r_{\rm max}\right)^{\alpha-2 }\right] +O\left(\beta ^2\right)\;,
    \end{eqnarray}
for the pressure, and
   \begin{eqnarray}
         \rho^{\rm UDM}(r) &\simeq& {v_{\max}^2 \over  r_{\rm max}^2}  \left[2 \beta \alpha { \left(1 +\alpha\right)  \over \left(2 + \alpha  \right)}   \left(r\over r_{\rm max}\right)^{\alpha-1 }\right] +O\left(\beta ^2\right)\;,
    \end{eqnarray}
using the same approximations and assumptions for the energy density.
In this case, it is clear that the NEC is preserved for both positive and negative $\beta$, when $|\beta|\ll 1$.\\

\begin{center}
\begin{figure}[h!]
\centering
    {\centering\Large\ \par\medskip}
        \includegraphics[width=11cm]{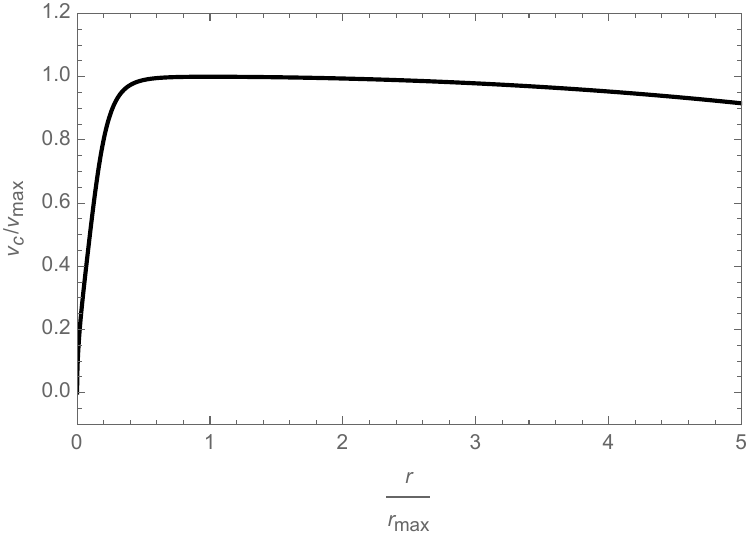}
    \caption{ Rotation curve, general case defined in Eq. (\ref{eq:general_case}), with $\alpha=2$, $\beta=0.01$ and $\delta=3$.   }
    \label{Fig3}
\end{figure}
\begin{figure}[h!]
\centering
    {\centering\Large\ \par\medskip}
        \includegraphics[width=\linewidth]{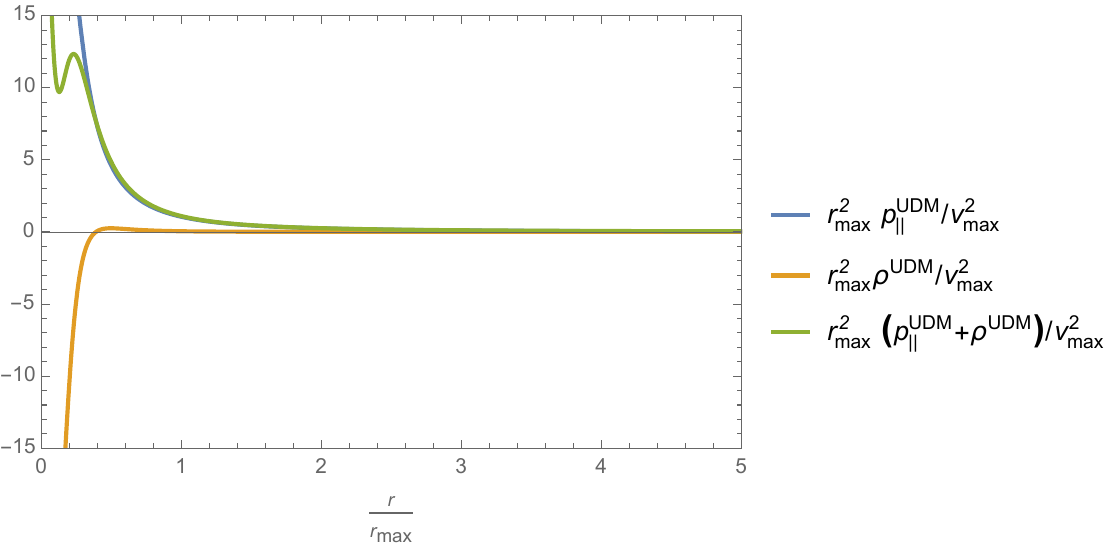}
    \caption{Energy density and pressure profile of a UDM model reconstructed from the rotation curve defined in Eq.~(\ref{eq:general_case}), with $\alpha=2$, $\beta=0.01$ and $\delta=3$.}
    \label{Fig4}
\end{figure}
\end{center}
Finally, let us conclude this subsection considering the general case 
defined in Eq.~(\ref{eq:general_case}) in which, e.g., we set $\alpha=2$, 
$\beta=0.01$ and $\delta=3$, see Fig.~\ref{Fig3}.
In Fig.~\ref{Fig4} we show $p^{\rm UDM}_\|$, $\rho^{\rm UDM}$ and 
$p^{\rm UDM}_\|+\rho^{\rm UDM}$ in the worst case, i.e., when we relax 
$r_{\rm min}=0$, $v_c(0)=0$ and $p^{\rm UDM}_\|(0)=0$. However, even 
in this limit, the NEC is satisfied.
In conclusion, starting from several possible parametrisations that can 
faithfully reconstruct the rotation velocity profile of a galaxy, we can 
always reconstruct a UDM model that satisfies the NEC even if in the 
bulge region near the centre we may have a negative energy density. 
However, as we emphasised above, we can always add a contribution 
$\propto 1/r^4$ to $p_\|$ and $\rho$ (related to $r_{\rm min}\neq 0$) 
that does not contribute to $v_c$ and that, at the same time, allows us 
to obtain an energy density with positive values [e.g., if $\beta >0$ 
and $p^{\rm UDM}_\| (r_{\rm min})>2 v_c^2(r_{\rm min})$ near the 
supermassive BH in the centre of the galaxy].

\section{Phenomenological analysis starting from CDM halo profiles} \label{sec:profiles}

In this section, we analyse different CDM density profiles presented in 
the literature. Some of the models discussed here are listed in the review 
\cite{Salucci:2018hqu}. As we shall see, for each CDM density profile, 
we are able to reconstruct a phenomenological profile of the energy density 
and pressure of our UDM model, using Eqs.~(\ref{eq:p_par-int_def-1}) 
and~(\ref{eq:rho-int_def-1}). The models we consider are the following.
We begin by considering the dark matter halo density defined by the 
empirical profile of Persic, Salucci \& Stel \cite{Persic1996}, the one 
obtained from N-body simulations \cite{Navarro_1996, Navarro_1997}, and 
finally the Burkert density profile \cite{Burkert_1995, Salucci:2000ps}.


\subsection{BT-URC}
\begin{center}
\begin{figure}[h!]
\centering
    {\centering\Large\ \par\medskip}
        \includegraphics[width=14cm]{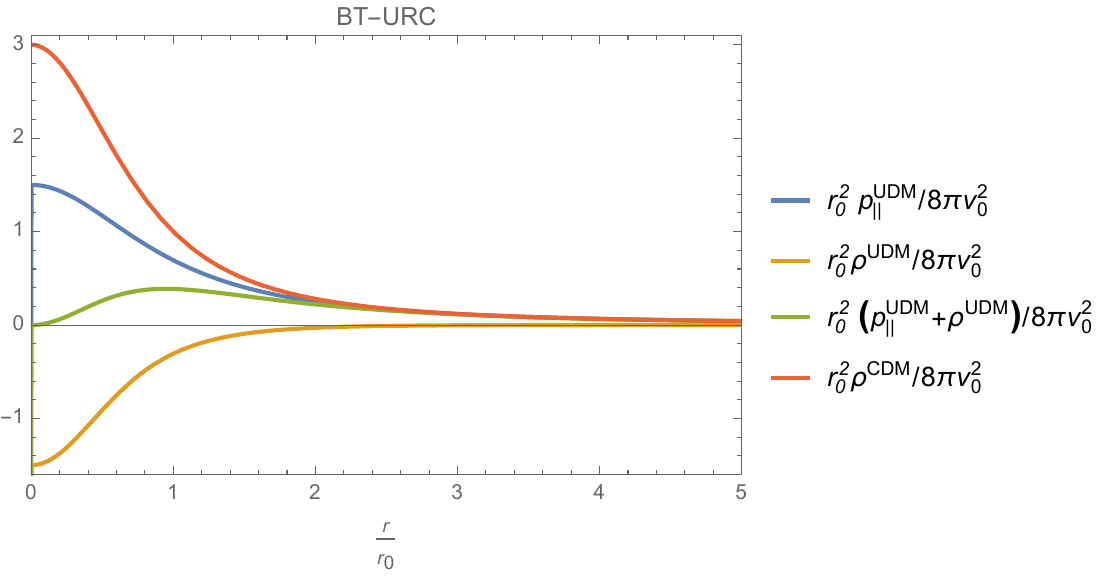}
    \caption{ Energy density and pressure profiles of a UDM model and matter density  profile of a CDM halo, reconstructed from the BT-URC rotation curve profile. }
    \label{Fig:BT-URC}
\end{figure}
\end{center}

This DM halo density profile was introduced in \cite{Persic1991, Persic1996} and 
takes the form (see also \cite{BinneyTremaine2008})
\begin{equation}
    \rho_{\rm BT-URC}^{\rm CDM}(r)={8 \pi v_0^2 \over r_0^2} 
    \frac{\left(3+r^2/r_0^2\right)}{\left(1+r^2/r_0^2\right)^2}\;.
\end{equation}
Here $r_0$ and $v_0$ are the core radius and the asymptotic rotation 
velocity of the DM halo, respectively. For the reconstructed UDM model, 
we find
\begin{eqnarray}
\label{eq:p-BT-URC_UDM}
  {p^{\rm UDM}_{\rm BT-URC}}_\|(r) &=& 
  {16 \pi v_0^2 r_0^2\over r^4} \Bigg[{r^2 \over r_0^2}+\frac{2}{1 + \left(r/r_0\right)^2}+\log \left({r^2\over r_0^2}+1\right)-{r_{\rm min}^2 \over r_0^2}-\frac{2}{\left(r_{\rm min}/r_0\right)^2+1}\nonumber\\
  && -\log \left({r_{\rm min}^2\over r_0^2}+1\right)\Bigg]  + {{p^{\rm UDM}_{\rm BT-URC}}_\|(r_{\rm min}) \over r^4} 
\end{eqnarray}
and
\begin{eqnarray}\label{eq:rho-BT-URC_UDM}
{\rho^{\rm UDM}_{\rm BT-URC}}(r) &=&  -{8 \pi v_0^2 \over r_0^2} \frac{\left(3+r^2/r_0^2\right)}{\left(1+r^2/r_0^2\right)^2} + {16 \pi v_0^2 r_0^2\over r^4} \Bigg[{r^2 \over r_0^2}+\frac{2}{1 + \left(r/r_0\right)^2}+\log \left({r^2\over r_0^2}+1\right)-{r_{\rm min}^2 \over r_0^2}
\nonumber \\ 
&& -\frac{2}{\left(r_{\rm min}/r_0\right)^2+1}-\log \left({r_{\rm min}^2\over r_0^2}+1\right)\Bigg] + {{p^{\rm UDM}_{\rm BT-URC}}_\|(r_{\rm min}) \over r^4} \;.
   \end{eqnarray}
   Now, relaxing the solutions where we set $r_{\rm min}=0$ and ${p^{\rm UDM}_{\rm BT-URC}}_\|(r_{\rm min}=0)=0$, as we see in Fig.\ \ref{Fig:BT-URC}, we are still satisfying the NEC.

\subsection{Navarro–Frenk–White-RC}

In the last decades, the development of increasingly accurate N-body 
simulations, describing the non-linear regime of structure formation, 
led to a constant improvement in the characterisation of galaxy density 
profiles. For example, the well-known Navarro-Frenk-White (NFW) profile 
\cite{Navarro_1996, Navarro_1997} and its generalisation \cite{White} are able to 
describe the universal form of the spherically averaged radial density 
profile of relaxed DM halos.
The NFW profile reads:
\begin{equation}
 \rho_{\rm NFW}^{\rm CDM}(r)= \frac{\rho_s}{(r/r_s)\left(1+r/r_s\right)^2},
\end{equation}
which contains two parameters that are strongly correlated and vary from 
halo to halo: the scale radius $r_s$ at which the slope of the profile 
changes from $1/r$ to $1/r^3$, and the characteristic density at $r_s$, 
which is $\rho_{\rm NFW}^{\rm CDM}(r_s)=\rho_s/4$. 
Following \cite{Navarro_1996} we can redefine $\rho_s=\rho_{\rm crit} 
\delta_c$, where $\rho_{\rm crit} =3H_0^2/8\pi G$ is the critical 
density, and $r_s= r_{200}/c$, where $r_{200}$ determines the mass of 
the halo $M_{200}=200 \rho_{\rm crit} (4\pi/3)r_{200}^3$. Here $\delta_c$ 
and $c$ are connected by imposing that the mean density within $r_{200}$ 
is $200 \rho_{\rm crit}$.

\begin{center}
\begin{figure}[h!]
\centering
    {\centering\Large\ \par\medskip}
        \includegraphics[width=14cm]{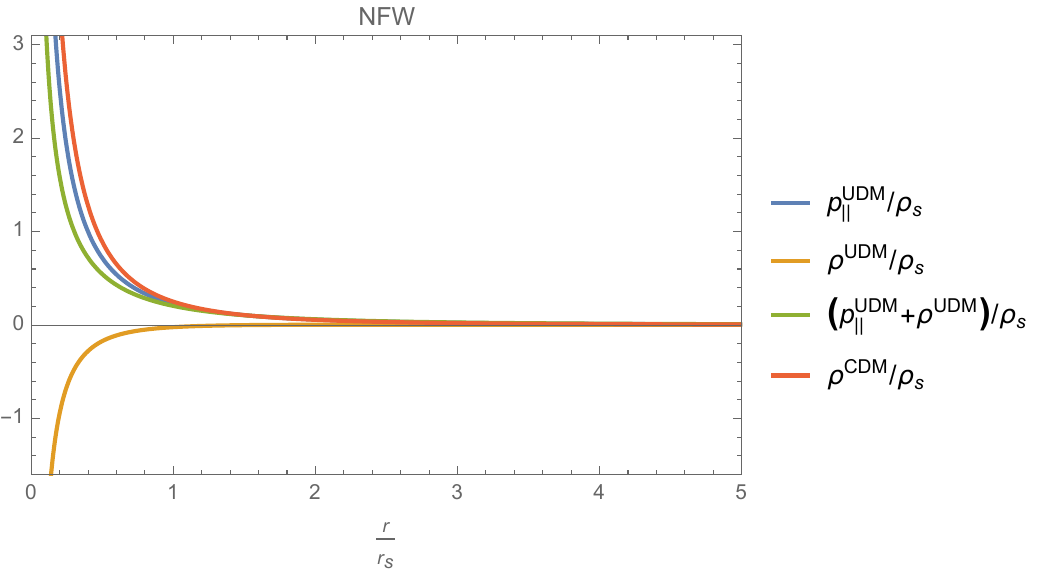}
    \caption{  Energy density and pressure profiles of a UDM model and matter density  profile of a CDM halo, reconstructed from the NFW halo profile curve profile.  }
    \label{Fig:NFW-UDM}
\end{figure}
\end{center}

Starting from NFW profile we can obtain for UDM models the following relations
\begin{eqnarray}
\label{eq:p-NFW_UDM}
  {p^{\rm UDM}_{\rm NFW}}_\|(r) &=&  {2\rho_s r_s^4 \over r^4 } \Bigg[ {r \over r_s}-\frac{1}{(r/r_s)+1}-2 \ln\left({r \over r_s}+1\right)-{r_{\rm min}\over r_s}+\frac{1}{(r_{\rm min}/r_s)+1}+2 \ln \left({r_{\rm min} \over r_s}+1\right)\Bigg]\nonumber\\
  && + {{p^{\rm UDM}_{\rm NFW}}_\|(r_{\rm min}) \over r^4} 
\end{eqnarray}
and
\begin{eqnarray}\label{eq:rho-NFW_UDM}
{\rho^{\rm UDM}_{\rm NFW}}(r) &=&  \rho_s \Bigg\{{2 r_s^4 \over r^4 } \Bigg[ {r \over r_s}-\frac{1}{(r/r_s)+1}-2 \ln\left({r \over r_s}+1\right)-{r_{\rm min}\over r_s}+\frac{1}{(r_{\rm min}/r_s)+1}+2 \ln \left({r_{\rm min} \over r_s}+1\right)\Bigg]\nonumber\\
\nonumber \\ 
&& - {r_s \over r \left(1 + r/r_s\right)^2}\Bigg\}+ {{p^{\rm UDM}_{\rm NFW}}_\|(r_{\rm min}) \over r^4} \;.
   \end{eqnarray}
   In Fig.\ref{Fig:NFW-UDM} we have assumed the worst scenario in which $r_{\rm min}=0$ and ${p^{\rm UDM}_{\rm NFW}}_\|(r_{\rm min}=0)=0$. It is also clear here that the NEC is satisfied.

\subsection{Bukert-URC}

\begin{center}
\begin{figure}[h!]
\centering
    {\centering\Large\ \par\medskip}
        \includegraphics[width=14cm]{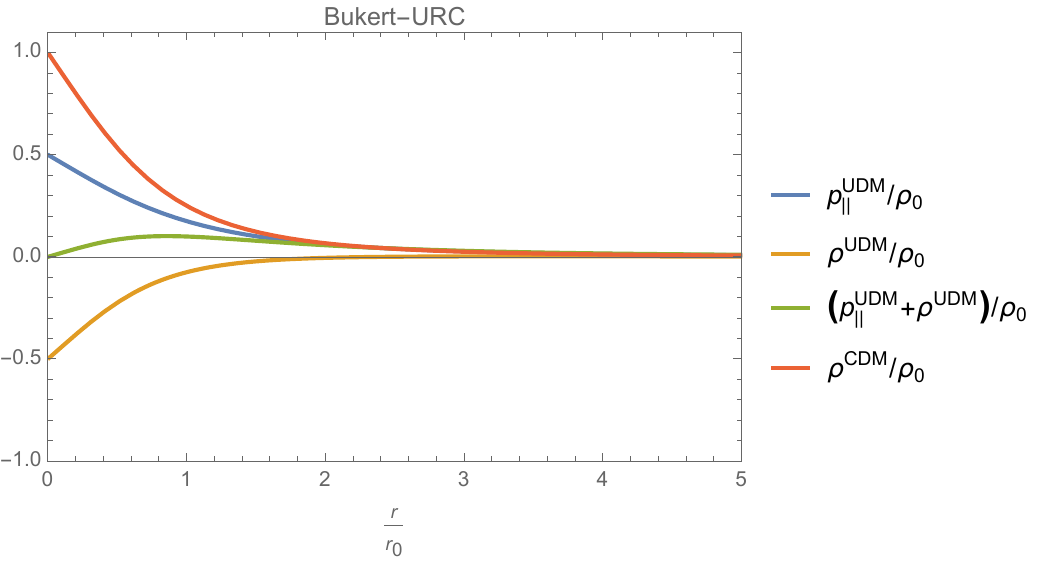}
    \caption{  Energy density and pressure profiles of a UDM model and matter density  profile of a CDM halo, reconstructed from the Bukert-URC rotation curve profile.  }
    \label{Fig:Bukert-URC-UDM}
\end{figure}
\end{center}

The Burkert density profile (B-URC), e.g.\ see 
\cite{Burkert_1995, Salucci:2000ps}, is an empirical profile that 
successfully matches the halo rotation curves of DM-dominated dwarf 
galaxies (for further analysis, see also 
\cite{Gentile_2004, Donato_2009, Salucci_2010}).
It is given by the following expression
\begin{equation} \label{B-URC_CDM}
 \rho_{\rm B-URC}^{\rm CDM}(r)= \frac{\rho_0 r_0^3}{(r+r_0)(r^2+r_0^2)}\;,
\end{equation}
where $\rho_0$ and $r_0$ are free parameters which represent the
central DM density and the scale radius \cite{Salucci:2000ps}.
For this profile we get
\begin{eqnarray}
\label{eq:p-B-URC_UDM}
  {p^{\rm UDM}_{\rm B-URC}}_\|(r) &=&  {2\rho_0 r_0^4 \over r^4 } \Bigg[ {r \over r_0} - {1 \over 2} \tan^{-1}\left( {r \over r_0}\right) %
  -{1 \over 2} \ln \left(1+ {r \over r_0}\right)
  - {1 \over 4}\ln \left(1+ {r^2\over r_0^2}\right) \nonumber\\
&& - {r_{\rm min}  \over r_0} + {1 \over 2} \tan^{-1}\left( {r_{\rm min} \over r_0}\right)
+{1 \over 2} \ln \left(1+ {r_{\rm min} \over r_0}\right) + {1 \over 4}\ln \left(1+ {r_{\rm min}^2 \over r_0^2}\right)\Bigg] \nonumber\\
 && + {{p^{\rm UDM}_{\rm B-URC}}_\|(r_{\rm min}) \over r^4} 
\end{eqnarray}
and
\begin{eqnarray}\label{eq:rho-B-URC_UDM}
{\rho^{\rm UDM}_{\rm B-URC}}(r) &=&  \rho_0 \Bigg\{{2 r_0^4 \over r^4 } \Bigg[  {r \over r_0} - {1 \over 2} \tan^{-1}\left( {r \over r_0}\right) %
  -{1 \over 2} \ln \left(1+ {r \over r_0}\right)
  - {1 \over 4}\ln \left(1+ {r^2\over r_0^2}\right) \nonumber\\
&& - {r_{\rm min}  \over r_0} + {1 \over 2} \tan^{-1}\left( {r_{\rm min} \over r_0}\right)
+{1 \over 2} \ln \left(1+ {r_{\rm min} \over r_0}\right) + {1 \over 4}\ln \left(1+ {r_{\rm min}^2 \over r_0^2}\right) \Bigg]
\nonumber \\ 
&& - \frac{ r_0^3}{(r+r_0)(r^2+r_0^2)} \Bigg\}+ {{p^{\rm UDM}_{\rm B-URC}}_\|(r_{\rm min}) \over r^4} \;.
   \end{eqnarray}
Also for this profile, for $r_{\rm min}=0$ and ${p^{\rm UDM}_{\rm NFW}}_\|(r_{\rm min}=0)=0$, we show in Fig. \ref{Fig:Bukert-URC-UDM}  the equation of state of a UDM model that has the same rotation curve of Bukert-URC profile. As we can see, also in this case NEC is totally satisfied.  

\section{Conclusions}\label{sec:conclusions}

In this work we investigated static, spherically symmetric halo configurations 
within Unified Dark Matter (UDM) scalar-field models and developed a systematic 
mapping between standard cold dark matter (CDM) profiles and their UDM 
counterparts, see also \cite{Bertacca-Halo_UDM}.
The starting point of our analysis, reviewed in Sec.~\ref{sec:UDMframework}, 
exploits the equivalence-class structure of UDM models: different Lagrangian 
realisations may share the same weak-field rotation curve while exhibiting 
distinct thermodynamic and field properties.
This allows us to disentangle kinematically fixed features from model-dependent 
ones within the UDM sector.

In Sec.~\ref{sec:rotationcurve}, we introduced a phenomenological description 
of the circular velocity profile, designed to capture the three characteristic 
regimes observed in real galaxies: a rising inner region, an approximately flat 
plateau, and the outer halo decline.
This kinematical input was  used in to 
reconstruct the effective energy density, radial and tangential pressures, and 
the scalar-field kinetic structure.
The presence of the free function $\lambda(r)$ was shown to be crucial: it 
parametrises the internal freedom of the equivalence class, enabling distinct 
UDM realisations to reproduce the same rotation curve while differing in 
microphysical properties.
At the same time, consistency conditions on the kinetic term ensure that the 
reconstructed models remain free from ghosts and instabilities throughout 
the halo.

In Sec.~\ref{Phenomenological_analysis-SC} we discussed how the energy density 
and pressure can be reproduced starting from an observed generic $v_c(r)$ of a 
spiral galaxy, both in the central bulge region and in the plateau/disk region. 
In this case, we also showed how the Null Energy Condition (NEC) can be preserved 
for a generic UDM halo.

Finally, we applied our reconstruction procedure in Sec.~\ref{sec:profiles} to 
several commonly used CDM halo profiles density models.
In all cases, we obtained UDM halo configurations capable of reproducing 
realistic rotation curves in the weak-field regime, demonstrating that the 
phenomenological success of CDM profiles can be retained within a relativistic 
UDM framework.
This result shows that the observed flatness of rotation curves can be reproduced 
in a unified description of the dark sector without introducing separate dark 
matter and dark energy components.

Overall, the formalism developed in this paper provides a unified approach to 
galactic dynamics in UDM models, clarifying which properties are constrained 
by observations and which remain flexible within the equivalence-class structure.
An explicit reconstruction of the UDM Lagrangians from the mapped 
profiles has not been addressed in this work and will be presented in a 
forthcoming paper.

\acknowledgments

DB and SM acknowledge support from the COSMOS network (www.cosmosnet.it) through ASI (Italian Space Agency) Grants 2016-24-H.0, 2016-24-H.1-2018 and 2020-9-HH.0.

\bibliographystyle{unsrt}
\bibliography{biblyosold}

\end{document}